\documentclass{acm_proc_article-sp}

\usepackage{algorithm}
\usepackage{algpseudocode}
\usepackage{paralist}
\usepackage{subfigure}
\usepackage{graphicx}
\usepackage{epstopdf}
\usepackage{epsfig}
\usepackage{color}
\usepackage{multirow}
\usepackage[hyphens]{url}
\usepackage{flushend}

\newcommand{\comment}[1]{}
\newcommand{\junk}[1]{}

\newlength{\figurewidthA}
\newlength{\figurewidthB}
\newlength{\figurewidthC}
\newlength{\figurewidthD}
\newlength{\figurewidthE} % 1 figure in a row (2 columns)
\newcommand{\ie}{{\em i.e., }}
\newcommand{\eg}{{\em e.g., }}

\newcommand{\advertisercount}{{3700 }}
\newcommand{\alexacount}{{1480 }}

\newcommand{\adscount}{{175,495 }}
\newcommand{\websitecount}{{180 }}
\newcommand{\profilecount}{{340 }}

\makeatletter
\def\@copyrightspace{\relax}
\makeatother

\renewcommand{\qed}{\rule{2mm}{2mm} \vskip \belowdisplayskip}

\begin{document}

\title{Adscape: Harvesting and Analyzing Online Display Ads}
\numberofauthors{3} 
\author{
\alignauthor
 Paul Barford\\
       \affaddr{University of Wisconsin-Madison}\\
       \email{pb@cs.wisc.edu}
\alignauthor
Igor Canadi\\
       \affaddr{University of Wisconsin-Madison}\\
       \email{canadi@cs.wisc.edu}
\alignauthor
Darja Krushevskaja\\
       \affaddr{Rutgers University}\\
       \email{darja@cs.rutgers.edu}
\alignauthor 
\and
Qiang Ma\\
       \affaddr{Rutgers University}\\
       \email{qma@cs.rutgers.edu}
\and
\alignauthor 
S. Muthukrishnan\\
       \affaddr{Rutgers University}\\
       \email{muthu@cs.rutgers.edu}
}
%\date{}
\maketitle
\begin{abstract}
Over the past decade, advertising has emerged as the primary source of revenue for many web sites and apps.  In this paper we report a first-of-its-kind study that seeks to broadly understand the features, mechanisms and dynamics of display advertising on the web - {\em i.e., the Adscape}.  Our study takes the perspective of users who are the targets of display ads shown on web sites.  We develop a  scalable crawling capability that enables us to gather the details of display ads including creatives and landing pages.  Our crawling strategy is focused on maximizing the number of unique ads harvested. Of critical importance to our study is the recognition that a user's profile ({\em i.e.,} browser profile and cookies) can have a significant impact on which ads are shown.  We deploy our crawler over a variety of websites and profiles and this yields over 175K distinct display ads.  

We find that while targeting is widely used, there remain many instances in which delivered ads do not depend on user profile; further, ads vary more over user profiles than over websites.  We also assess the population of advertisers seen and identify over 3.7K distinct entities from a variety of business segments.  Finally, we find that when targeting is used, the specific types of ads delivered generally correspond with the details of user profiles, and also on users' patterns of visit. 
\end{abstract}

\section{Introduction}
\label{sec:intro}

Advertising online is a compelling proposition for brands and e-commerce vendors that seek engagement with a broad cross-section of potential customers.  The sheer volume of online users and the increasing amount of time that people spend online has led to an estimated \$36B in online ad spending in the US for FY2012, which represents an 15\% increase over the previous year~\cite{iab12}.  The majority of this spending is on advertising that most commonly appears in search results as text ads.  There is, however, a growing preference for display ads --- typically image and video ads that appear in response to users' browsing and other activities on the web --- that can convey more robust and visual messages to users.  A recent report by Forrester estimates that \$12.7B was spent in the US on display and video advertising in FY2012, and is growing at 17\% annually~\cite{Forrester}. 

The ubiquity of advertising on publisher web sites and apps makes it easy to overlook the diversity and complexity of the online ad delivery ecosystem. Well known depiction of the online ad ecosystem is the {\em Display Lumascape} that shows Marketers (brands) and Publishers/Consumers connected by hundreds of companies that provide a variety of intermediary services~\cite{lumascape}.   The task of delivering billions of ads from thousands of Marketers to millions of Publishers' web sites and apps on a daily basis is quite daunting, hence the complexity of the system.

Of central importance in the ad delivery process is the selection of a specific ad to display to a given visitor of a Publisher site.  This is generally referred to as the {\em targeting problem}.  Targeting is commonly based on criteria like site/page context, placement size, user behavior and geolocation.  Ad serving infrastructures provide targeting information that enables Marketers to bid on specific ad requests from a large number of Publishers.  The presumption is that improvements in ad targeting will benefit all of the constituents in the ad ecosystem.

In this paper, we present a first-of-its-kind study of display advertisements that are being delivered to online Consumers.  Our objectives are to broadly characterize the online display advertising landscape or {\em Adscape} and to elucidate targeting mechanisms from an empirical perspective.  The study seeks to better understand the range of online ads, the degree of similarity between ads shown to different visitors of the same web site, the breadth of ads that are shown on a given web site, and the degree to which ads shown on different web sites are different from each other.  In the long term, we hope to provide a foundation for improving ad targeting mechanisms and streamlining the ad-serving ecosystem.

The capability to gather display ads from across the web is central to our work~\footnote{While display ads are also delivered to apps, we restrict our focus to the web for this study.}.   The vast number of web sites that run display ads, the diversity and dynamics of the targeting mechanisms, and the need to minimize the impact (load due to measurement) on any given web site all present significant challenges.  We addressed these challenges by developing  
methodologies, tools and infrastructure for ad centric web-crawling.  At the heart of our system is the notion of {\em profile-based crawling}, that enables each crawler instance to interact with the ad ecosystem as though it were a unique user with specific characteristics.  This capability is essential for collecting the full spectrum of ads that are delivered across web sites.

This crawling infrastructure was used to gather over 175K distinct ads and associated data (landing pages, creatives, etc.) from \websitecount large
 English language web sites that run display ads~\footnote{While we restrict our focus to English language web sites for this study, our methodology and tools can be more generally applied.}.  We use 340 different user profiles in our crawling experiments.  We make no claims that this sampling of web sites shows {\em all} available (English) display ads since local advertisers often focus on smaller, local sites.  We employed different crawling strategies to assemble a data corpus that is extensive enough to expose key characteristics of the Adscape and provide insights on the most commonly-used targeting mechanisms.  

When analyzing our data, we first consider the general targeting mechanisms that are used on different sites.  Not surprisingly, we find that the majority of sites do indeed use targeting mechanisms on over 80\% of their ad inventory.  However, many sites show a substantial number of ads to all users regardless of profile.  We next consider the Marketers who are engaged in online display advertising.  Our analysis shows over 3.7K distinct Marketers from diverse business segments such as shopping, computer sales and financial services.  Third, we drill down on the details of the ads themselves.  Our analysis reveals that  
\begin{inparaenum}[(i)]
\item interest-based targeting, which attempts to ensure that ad types shown generally align with the customer's interest profile characteristics, is wide spread, and
\item age and gender-based targeting is also widely used.
\end{inparaenum}

In summary, we have developed profile-based web crawling capabilities and strategies that enable diverse display advertisements to be gathered from a large set of web sites.  We have applied this crawler to a selection of large English language web sites to gather a sizable corpus of display ads~\footnote{We plan to open source our ad crawler and make this data corpus available to the community upon publication.}. In this study, we evaluate and characterize this data and report its characteristics from a variety of perspectives. 

\section{The Anatomy of Display Ads}
The earliest online display ads, which appeared in the mid-1990's, were simply a bright, flashy billboard that was shown to {\em any} visitor to the page over some period of time.  The technology for serving ads has evolved tremendously since then, and now ads are typically  \emph{targeted} to users. %(\eg Figure~\ref{fig:first_ad}).  
We can look at the diversity of modern display ads from several different perspectives.

\medskip
\noindent
{\bf Advertiser View.}
Whether branding a product or drawing attention to a special discount, an advertiser wishes to attract the attention of potential customers.  They design campaigns to achieve this goal.  Campaigns include creatives, target demographics, frequencies, placements, etc.  All of these considerations must be balanced with an advertising budget.  Two example ad campaigns are as follows: 
\begin{inparaenum}[(1)]
 \item  \emph{A company that produces running shoes targets users who live in Madison, WI, from 6AM to 9AM on business days, and the ad must not be shown to the same user more than 7 times on any given day.}  This campaign targets users based on location (``geo'' in ads parlance), time of the day (one or more ``dayparts''), and further limits number of times the ad may be shown in a period of time (``frequency cap''); 
 \item \emph{A company that produces running shoes targets male users aged $[25-30]$ who are interested in ``sports", ``healthy lifestyle"  and ``jogging".} This campaign targets users based on their interests (``profile'' in ads parlance) and also demographics (gender, age).
 \end{inparaenum}

Targeting strategies can be combined in sophisticated ways by advertisers, and the industry relies on existence of players who can track cookies and maintain user profiles. 

\medskip
\noindent
{\bf Publisher View.}
A Publisher produces content or provides services that attract users, and with that, the opportunity to present ads to those users. On any given visit, ads can be served from a variety of sources including 
\begin{inparaenum}[(i)]
\item {\em premium campaigns}, which are contracts with specific advertisers, 
\item {\em ad networks}, which represent multiple advertisers, and 
\item {\em ad exchanges}, which offer an auction-based environment for matching publishers with advertisers.
\end{inparaenum}

Typically, publishers combine the methods, even on a single page.

\medskip
\noindent
{\bf The Adscape View.}
Consider a user $u(t)$ accessing a webpage $w(t)$ at time $t$.  Say the publisher of $w(t)$ shows a set of ads $a(t)$ to  $u(t$). There is some {\em allocation function} $f_w(t): u(t) \rightarrow a(t)$. The set of all $f_w(t)$'s over all $w$'s and all users $u(t)$ at any time $t$ will be the {\em Adscape} that is the focus of this paper. 
The functions $f_w(t)$'s may depend on:  
\begin{inparaenum}[(i)]
\item user's  demographics, interests, location, etc; 
\item site $w$, its contents and context; 
\item time $t$, and the past, including users' past actions, $w$'s past contents, while  $f$ may vary over time; 
\item the set of ads $a()$'s, including mutual constraints that allow or disallow each other; 
\item
mechanisms, incentives and market conditions that govern the behavior of advertisers, networks, exchanges and publishers.
\end{inparaenum}
The function $f_w(t)$ may ultimately be simple (all users see the same ads for a day) or sophisticated (each ad is personalized based on multiple criteria). 
Our research agenda is to broadly understand the user-targeting aspect of the Adscape of online display advertising. We pursue this goal by crawling $w$'s, detecting and harvesting ads $a()$'s, and thereby observing and finding patterns in $f_w(t)$'s. 

\section{Challenges in Observing the Adscape}
\label{sec:challeges}
In general web crawling, the goal is to identify and possibly index the {\em contents} (ads aside) of all the webpages $w$ that are online. In what follows, we compare that with the problem of crawling and understanding $f_w(t)$'s and ultimately the Adscape.

\medskip\noindent {\bf Web Content vs. Ads.}
Web  crawling is challenging because the existence of a page $w$ may be unknown. Our problem of crawling for $f_w(t)$'s has similar challenges in discovering $w$'s. However, there are additional challenges.  For example, given some page $w$, 
it is nontrivial even to identify which elements are ads, and which are regular content. Further, given an element that {\em is} an ad, it is typically not represented in a HTML code, but in some other form (usually Javascript), and several executions may be needed to actually retrieve the ad.

\medskip\noindent {\bf Dynamic content and recrawl rate.}
Web crawling is challenging because $w$ may have dynamic content that varies over time. To address this, robust web crawling methods have been developed, including the ability to identify content change frequency and using that to specify recrawling rates.  In contrast, ads are far more dynamic --- a popular web page may show many different of ads depending on a user's geographic location to another, the time of day, etc. even if the content on the page does not change.  This complicates calibration of ad recrawling rates. 

\medskip\noindent {\bf Personalization.}
Arguably some web content is personalized to the viewer, thus web crawlers have to mimic viewers to collect this content. In many cases, the personalized content may not even be relevant to crawl.  In contrast, ads, are critically targeted to users' profiles, and it is imperative to mimic multiple user profiles in order to understand the Adscape.

\medskip\noindent {\bf Contamination.}
Ad crawlers that visit pages end up contributing to data observers on the web who build profiles of users based on the browsing behavior. Therefore, even visiting a page modifies the profile associated with the crawling session, which can contaminate the profile that a crawler adopts. Further, intermediaries like the ad networks observe browser sessions and adopt their strategies, leading to potentially further contamination.

\medskip\noindent {\bf Economics.} 
Web crawlers should be calibrated not to overload the sites they crawl. Similar consideration holds for ad crawlers, but there is a deeper concern. Display ads are charged {\em per impression}.  Thus, each time the crawler accesses a page and associated ads, advertisers incur a cost. Therefore, an ad crawler must be configured to limit costs for the advertisers, which in turn limits the sample of the space that is observed. 

\medskip
Thus ultimately, an ad crawler needs to 
\begin{inparaenum}[(i)]
\item search over far more states than a corresponding web crawler, proportional not only to the number of pages, but also the number of user profiles, geo locations, day parts, etc. 
\item minimally distort the statistics of ads being displayed (and hence have minimal cost to the advertisers) 
\item and prevent contamination of a crawling profile. 
\end{inparaenum}
 According to WorldWideWebSize\footnote{\url{http://www.worldwidewebsize.com/}}, as of May 03, 2012, the Indexed Web contains at least 14.24 billion pages. The state an ad crawler has to explore is a multiple of this number. If we assume 1000's of geos, 10's of daily segments, 1000's of profiles, this altogether yields multiplier in the range of $10^7$ or more!   Ultimately one must make some simplifying assumptions to shrink this state space. 

Finally, one could ask if the Adscape can be studied without crawling?  In the abstract this could be done if one could partner with all advertisers, but this would seem to be an infeasible task.  It would also preclude understanding the targeting details of ad delivery.  Similar challenges exist with possible partnerships with intermediaries or publishers due to simple issues of scale.  Thus, we focus on crawling as our approach to data gathering and consider ways in which we can address the state reduction issue toward the goal of gathering a representative sample of the Adscape.

\section{Our Approach}

In this work we focus on the impact of user interest-based ad personalization or \emph{user interest-based Adscape}.  More formally, we restrict $f_w(t): u(t) \rightarrow a(t)$ as follows:  
\begin{inparaenum}[(1)]
\item we fix geolocation (by performing all data collection from a single location); 
\item we do not consider time-of-day effects.  
\end{inparaenum}
Both dimensions are important and will be the subject of future work.  This paper focus is on $f_w(t): p(u(t)) \rightarrow a(t)$ where $p(u(t))$ is the {\em profile} (or {\em persona} which we use interchangeably) of the user. We will make $p()$ more precise in the future, but it encompasses users' interests. We refer to $(w, p)$ as a \emph{pair} where $w$ is a website and $p$ is a persona. 
While distribution of different types of personas in Internet can be arbitrary and can be a parameter of $f_w(t)$, in this paper we restrict our attention to targeting algorithms given a $(w, p)$ pair. In the future, one could expand the study by exploring actual frequencies of different user types and shape of the traffic (\eg using comScore).

Let $\mathcal{W}= (w_1, w_2, \dots)$ be set of all websites, and $\mathcal{P}=(p_1, p_2, \dots)$  be a set of all personas. In general, we will not be able to use all pairs formed from $\mathcal{W}$ and $\mathcal{P}$ for crawling because of the imposed load on our systems as well as ad ecosystem. Hence we approach it in four steps:
\begin{inparaenum}[(1)]
\item Given a single pair, $(w,p)$, we crawl the pair --- crawl site $w$ with browser depicting persona $p$ -- several times in a row,  study the distribution of ads over time, and propose a  pattern of crawls ({\em crawling strategy}) we will ultimately deploy  for the pair.
\item Create  a large pool of websites $W\subset \mathcal{W}$ and pool of personas $P \subset \mathcal{P}$ which will be basis for our research.
\item Crawl all possible pairs formed from $W$ and $P$ for a short period of time. Analyze the data to identify a small {\em focus set}: a subset of pairs that we will crawl operationally. The choice is done with a budget in terms of the number of crawls we can do with the crawling strategy above. 
\item We crawl  the focus set of pairs as per the strategy, log the crawls and collect data about the ads observed.  
\end{inparaenum}
Details of these steps are described below. 

\medskip\noindent\textbf{Crawling Strategy.}
\label{subsec:crawlingstrategy}
Intuitively, we want to maximize total number of distinct ads. Our expectation is that the large corpus of distinct ads will allow us to detect targeting patterns.  We are not aware of any previous work that discusses this problem. 

Ads shown to a pair $(w, p)$ can be observed in many different ways. For instance, we can visit $w$ and collect ads once every hour for a week, or we can visit $w$ many times in rapid succession. Results will likely be different.  For instance, if the user makes too many visits to a single page, she can be classified as a ``bot", and as a result might only see the limited selection of ads.  Alternatively, if advertisers use frequency caps and the cap is small, then sequential visits of the website with $p$ can yield many distinct ads. If frequency capping is used rarely, then crawling may collect only a small selection of ads. Hence, to maximize the number of distinct ads, we will pursue 2 strategies: 
\begin{inparaenum} [(1)]
\item short and
\item long. 
\end{inparaenum}
The strategies are characterized by two numbers: $\alpha$ --- number of rapid sequential visits, and $\beta$ --- number of repetitions. Note, that initial information about $p$ before each repetition is identical.

\medskip\noindent\textbf{Website Pool $W$.}
The number of distinct webpages on Internet is huge, and grows each day.  However, in practice only few of these are frequently visited.  For this study, we create a pool $W$ from top popular websites using Alexa\footnote{\url{http://www.alexa.com/topsites}}.

\vspace{2mm}\noindent\textbf{Persona Pool $P$.}
We model a persona as set of user profile interests that are associated with the user (or her browser, to be more precise). We build $P$ from Google's advertising interests tree.  For this study, we choose interest categories such that $P$ is diverse and represents interests that are popular in Internet.

\vspace{2mm}\noindent\textbf{Selecting a Focus Set.}
Let $S=\{(w_i, p_j)\}$ be set of all possible pairs, such that $w_i \in W$ and $p_j\in P$.  We want to select focus set $C\subseteq S$, such that the size of $C$ is at most $B$, where $B$ is the budget in number of pairs.  $B$ takes into account crawling strategies, as well as network, server and bandwidth constraints.  To maximize the total number of distinct ads, we observe $S$ for a limited time and choose $C$ of size $\le B$ that produces a large number of distinct ads.

It is easy to see that this is a budgeted maximum cover problem. Let $\{a_{ij}\}$ be a set of ads displayed on website $w_i$ for profile $p_j$. Given a set $A = \cup_{(i,j)}\{a_{ij}\}$, we want to choose subset $C\subset S$, s.t., $|C|\le B$ and $\forall T\subset S$, s.t., $|T| \le B$ the following holds:  $|\cup_{(w_i, p_j)\in C}\{a_{ij}\}| \ge |\cup_{(w_i, p_j)\in T}\{a_{ij}\}|$.
This problem is $NP$-hard, hence we proceed with approximate greedy solution that is known to be an $1-\frac{1}{e}$ approximation to the optimal. The description of greedy algorithm can be found in Algorithm~\ref{alg:max_cover}.  
%The idea of the algorithm is as follows: take all pairs $\{w_i, p_j\}$ and associated with them ads $a_{ij}$. 
We start with empty cover $cover$. We sort pairs in decreasing order of number of distinct ads ($o$), ties are broken arbitrarily. While the budget is not exhausted, take the first element of the ordering $h$, remove it from the list, and add to $cover$. Process remaining pairs in $o$ by removing ads $u$ that occurred for pair $h$.  Keep adding element to $cover$ until the budget is exhausted or ordering $o$ is empty. Output $cover$.

\begin{algorithm}
\caption{Select Subset of Pairs}
\label{alg:max_cover}
\begin{algorithmic}[1]
\Procedure{FindMaxCover}{$\{w_i, p_j, a_{ij}\}, B$}
\State $cover \leftarrow []$, $u \leftarrow []$
\State $o \leftarrow$ \textit{pairsInDecrOrderOfNumOfAds}$(\{w_i, p_j, a_{ij}\})$
\For{$i \in \{1,\dots, B\}$}
	\State $h \leftarrow o.pop()$
	\State $cover \leftarrow cover \cup h.getPair()$
	\State $u \leftarrow u \cup h.getAds()$
	\State $removeAds(o, u)$
	\State $o \leftarrow$ \textit{pairsInDecrOrderOfNumOfAds}$(o)$
\EndFor
\State \textbf{return} $cover$ 
\EndProcedure
\end{algorithmic}
\end{algorithm}

\section{Profile-Based Crawling}

Our overall profile-based crawling system uses novel methods for 
profile generation, uncontaminated profile crawling, and ads collection and classification. In this section, we describe these methods.

\subsection{Profiles}
\label{subsec:profilebuilder}
\noindent
\textbf{Profile Generation.}
Profiles are generated for users based on their interactions with websites. 
Amongst the common browser cookies and non-cookies based (\eg Flash cookies, server side profiling) web tracking technologies, 
ad networks and ad exchanges typically associate user properties with users via browser cookies. 
This basic observation implies a mechanism for building profiles.

There are several ways to create user profile. The first approach is to mimic actions of a particular user (\eg ~\cite{castelluccia2012betrayed, Liu11}). Characteristics of profiles built in this fashion can be arbitrary close to the profiles observed in reality. However, the approach is impractical for the exploration of the large space of profiles.  A second approach is to establish many versatile profiles by going to webpages of ad networks and establishing profiles manually\footnote{\eg \url{www.google.com/settings/ads}}. However, using this approach, only the ad networks for which the profile was set, will recognize the user. Given the complexity of the ecosystem, this approach is likely to be insufficient.  A third approach is to create \emph{user interest-based} profiles by crawling a set of target sites. This approach can potentially generate profiles for {\em any} user interest depending on the sites visited.  It assumes that a user is assigned to profile categories based on websites visited and is presented on ad servers as a ``bag of interests".  

We selected the crawler-based approach to explore targeted ad serving broadly. We have implemented Profile Builder as follows. For a given interest category, Profile Builder 
\begin{inparaenum}[(1)]
\item fetches the top 50 websites associated for this category (\eg using Adwords Ad Planner); 
\item opens Firefox with an empty profile (\eg using  Selenium WebDriver\footnote{\url{http://docs.seleniumhq.org/projects/webdriver/}}) and visits fetched URLs;
\item visits Google ads settings editor page and captures it's content. This step is not required for profile generation, however it allows us to check the interest categories assigned to the profile;
\item zips the profile, including the cookies;
\item stores the profile for future use.  
\end{inparaenum} 
One can verify the \emph{state of the profile}, or interest  categories assigned to it, at any given moment, by opening Firefox with the profile and visiting the corresponding pages of ad networks.
There are multiple sources for input interest categories. For instance, interest trees used by ad networks to allow advertisers to choose their target audiences.

This approach has several limitations. Profiles treated in this way do not take into account server side profiling or Flash cookies. Another limitation is in the ability to discover \emph{retargeted} ads, since in order for profile to get retargeted it has to visit the website of the advertiser first.  However, retargeted advertising, browser and device-based targeting differences are out of the scope of this study.

\medskip
\noindent
\textbf{Initial Profile Contamination.}
As discussed earlier, for each target interest category Profile Builder uses a list of websites. In general websites have more than one category, for instance Google Display Planner associates categories \textit{News/Business News, Finance/Investing} with \url{bestonlinetrading.info}. Intuitively, categories describing content  are assigned to a user profile. Hence, the generated profile can have multiple categories, in addition to target categoty, associated with it. 
For instance, the profile created for interest category \textit{Finance/Accounting \& Auditing} contains interests \textit{Business \& Industrial}, \textit{News}, and \textit{Computers \& Electronics} etc. 

\begin{figure}[t!]
 \centering
  \psfig{file=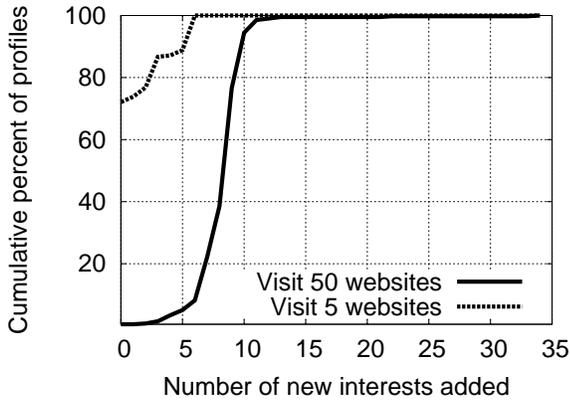, width=\figurewidthC, angle=-90}
\caption{Dynamics of the number of profile interests as more websites are visited.}
\label{fig:contamination}
\vspace{-0.1in}
\end{figure}

\medskip
\noindent
\textbf{Uncontaminated Crawling.}
Profiles are dynamic and change as more websites are visited  (\eg new categories are added). For the purpose of our experiments,  we refer to this phenomenon as \emph{profile contamination}. Profile change is undesirable when we seek to understand how ads are targeted at a particular profile. To get a better understanding of this phenomena we conducted the following experiment. We sampled 60 profiles from $P$, and crawled 50 random websites from $W$.  We checked the state of the profiles using Google's ads settings editor page after each visit.  Figure~\ref{fig:contamination} shows the distribution over the number of new categories obtained after visiting 50 websites. The figure shows that 60\% of profiles gained more than 9 new interests after 50 visits. That is significant, as average number of interests in the beginning of the  experiment was 8.  However,  this issue can be mitigated by limiting website visits to 5 or even fewer.

\vspace{-0.1in}

\subsection{Harvesting and Identifying Ads}
\noindent
\textbf{Harvesting ads.}
Ads are often delivered to webpages by JavaScripts that are executed at the time of page load. For instance, it is not sufficient to simply send a HTTP request and parse the response, since the ads will not be initialized.  To address this, one could potentially call JavaScripts.  However, this requires calling ``proper" JavaScripts with ``proper" parameters.  Alternatively, one can let browser do the work and see the page exactly the the way a user would see it.  We chose to stay agnostic to JavaScript execution, and load the pages directly in a browser.

We have implemented a Firefox extension \emph{Firefly}, that is able to control the instance of Firefox. For example, Firefly can receive a command to load $w$.  Once $w$ is fully loaded, including all iFrames and JavaScripts, Firefly parses the source of $w$ and harvests \emph{visual elements}.  Visual elements are objects appearing in \texttt{img} and \texttt{embed} HTML tags. Most of the display ads are shown using these two tags.  For each visual element Firefly captures: 
\begin{inparaenum}[(1)]
\item the source of the element; 
\item the URL of an iFrame the element appears in (if any); 
\item the landing page - URL where user is directed in response to a click; 
\item the dimensions of the element (width and height).
\end{inparaenum}

\medskip
\noindent
\textbf{Identifying Ads.}
It is sometimes hard to find ads on a given webpage, even for a human.  We have developed a three step decision process, that identifies ads automatically.  In order to be classified as an ``ad" a target visual element has to pass following tests:  
\begin{inparaenum}[(1)]
\item {\em AdBlock Test}. Adblock's easy list \footnote{\url{http://easylist.adblockplus.org}}  is a database of regular expressions that can be used to detect ads.  We test an element's URL, iFrame URL, \texttt{div} class and a landing page URL against it to see if a match can be found; 
\item {\em Dimension Test}.  Display ads frequently have standard sizes (\eg a standard banner is 728$\times$90), which enable them to fit into design of many pages.  We maintain a list of 25 different standard dimensions. The visual element has to match one of the entries of the list; 
\item {\em Self Ads}. A visual element cannot link to a page within the same domain. Ad elements have to have an external link.  
\end{inparaenum}
If visual element passes all tests, it is classified as ``ad".
 
\vspace{2mm}
\noindent
\textbf{Parsing the Landing Page of the Ad.}
The landing pages, as given in HTML of visual elements, do not necessarily directly link to an advertiser's site. Indeed, there are often multiple redirects before ``settling". Redirects can be used for accounting purposes (\eg click counting). For our analysis, we are interested only in the {\em final} destination in the chain. Our Parser fetches the URL using a simple HTTP library, and receives the destination URL, which is stored as a landing page of a visual element. We attempt to decode original landing page URL's query string, if an HTTP request gets blocked. 

\subsection{Overall Solution}

We have implemented a distributed ad harvesting and parsing infrastructure based on the methods described above.  Ad harvesting is managed by the \emph{Controller}. The Controller is configured with a \emph{crawling plan} --- a list of persona/site pairs to crawl with specified frequency --- as an input, and executes it while balancing the load.  The Controller manages the number of Firefox instances (also stated in the crawling plan) that are administered via our Firefly extensions. Most importantly, the Controller can open Firefox with profile $p$ as per the crawling plan. The Controller sends commands to Fireflies to visit $w$'s respecting profiles $p$ that the Firefox's are using.  The Fireflies follow the order and report back harvested visual elements, that are stored into database. 

Parser instances function independently of the harvesters. They take unprocessed entries populated by harvesting, and parse them.   In addition they (1) identify ads and (2) resolve ad landing pages. The Parser also downloads all visual elements and stores a local copy of them.
\section{Data Collection}
\noindent
\textbf{Selecting $W$ and $P$.}
For initial pool of websites $W$, we made a selection of popular websites from Alexa\footnote{\url{http://www.alexa.com/topsites}}.  We took the top $1500$, removed non-English websites, websites with adult content and sites that contain no ads.  The result  was {\bf 314 websites}.  

We used Google's advertising interests tree as a basis for creating the initial pool of personas $P$.  We chose it because Google is the leader in display advertising, and has the largest user interest category tree that we could find. To diversify the pool, we picked all second level interests (251).  Furthermore, to ensure that $P$  included interests popular in Internet, we proceeded as follows: 
\begin{inparaenum}[(1)]
\item select the top 1,000 websites from Alexa that are part of Google Display Network;  
\item for each website in the list, we used Google Ad Planner \footnote{\url{https://www.google.com/adplanner/}} to get the 10 most popular interests that were present in profiles of users who visit the website;
\item we formed the list of interests that were captured, and then added all third level interest categories found in the list to $P$.  
\end{inparaenum}
The result of this process was 340 interest categories, and a total of {\bf 340 profiles} are created corresponding to each of them.

\vspace{2mm}
\noindent
\textbf{Initial Contamination of Profiles.} 
To measure initial profile contamination, we observed the initial state of generated profiles. We collected interest categories assigned to each distinct generated profile by visiting the Google ads settings editor page with Firefox opened using the profile and calculated the fraction of \emph{relevant} categories. We call a category \emph{relevant}
if it is located in the subtree rooted at the category for which the profile was created, otherwise we refer to it as \emph{irrelevant}.  

The vast majority of profiles contain a significant number of irrelevant categories. We have also observed that profiles created for Health-related topics frequently have 0 relevant interests assigned to them. Exceptions are  \emph{Health/Aging \& Geriatrics}, \emph{Health/Pediatrics} and \emph{Health/Women's Health}. This phenomenon can be explained by the fact that health-related categories are more sensitive than others. All of the profiles generated for $3^{rd}$ level categories have less than 20\% relevant interests. This is expected, for instance because even the direct parent of the $3^{rd}$ level category is counted as irrelevant, according to our definition. Finally, we did not observe any $1^{st}$ level category whose $2^{nd}$ or $3^{rd}$ level profile categories would be consistently getting high or low relevance fractions.

\begin{figure}
\centering
\includegraphics[width=2.3in]{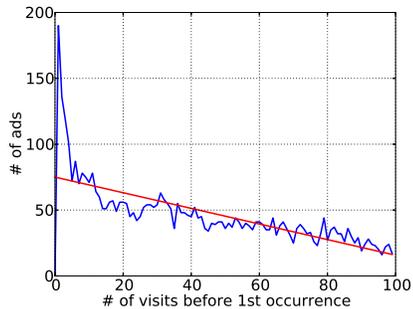}
 \caption{Arrival of new ads (average rate).}
\label{fig:xyz}
\vspace{-0.1in}
\end{figure}

\vspace{2mm}
\noindent
\textbf{Crawling strategies.}
We began by selecting 100 random pairs formed using $W$ and $P$ and visiting each sequentially 100 times (total of 10K visits).  We observed  $\approx 3K$ unique ads.  We studied the average arrival rate of new ads (see  Figure~\ref{fig:xyz}). We observed that in the beginning of the session a given profile was shown many new, yet unobserved by it, ads. At approximately visit 5 the arrival rate of the new ads drops drastically. We posit that this threshold approximates the average duration of a user session on the websites.  We also found that beyond the tenth visit, the rate at which new ads were served slowly decreases. This roughly follows a linear function $y=-0.6x +75$ (red line on the plot). Using this insight, we define our {\em short} strategy by $(\alpha, \beta) =(10, 5)$.  The intuition for this is that an average user session on the sites in $W$ is 3.7, and exceeds 10 for only 6 websites.  For the {\em long} strategy we choose parameters $(\alpha, \beta) =(100, 1)$, to explore the effect of the larger number of visits.  We decided to keep the length at 100, because during the experiment described above none of our machines was identified as ``bot" in the sense that public service ads were shown or ads stopped altogether.

\vspace{-0.1in}

\vspace{2mm}
\noindent
\textbf{Focus Set Selection.}
\label{sec:focus_set}
We set a budget to approximately 300K visits for the actual data collection (excluding profile generation). This results in a focus set of  $\approx 1900$ pairs ($1900 \times 100 + 1900 \times 10 \times 5 =  285K$).
To form the focus set we first form all possible $314\times340$ pairs. We visit each pair 5 times, to limit profile contamination, as discussed in~\ref{subsec:profilebuilder}.
Using the collected data, we first built a budgeted max cover as described in Algorithm~\ref{alg:max_cover} with budget set to 1M. 
We formed the final focus set as follows:  
\begin{inparaenum}[(1)]
\item we took top 1700 pairs from the cover; 
\item we want each $p$ and $w$ to be present in at least 3 pairs. To achieve that we traversed the cover ordering starting from position 1700, and added required pairs, totaling in 79.
\item for each distinct website $w$, we added a pair $(w, e)$, where $e$ is an \emph{empty} profile ({\em i.e.,} a profile that was not used for any browsing and hence has no interest categories associated with it), for the baseline. 
\end{inparaenum}
The resulting focus set was 1959 distinct pairs. These pairs contain 180 distinct websites and 340 distinct profiles {\em i.e.,} we  selected \textit{all} $p \in P$.

\vspace{2mm}
\noindent
\textbf{Dataset.}
Finally, we used our crawler to collect data on the focus set.  Data was collected over a two day period from 10/1/2013 to 10/3/2013.  This data collection produced 875,209 impressions, and 175,495 distinct ads.  We observed ads from 3,700 advertisers served using 106 distinct ad servers.

\vspace{2mm}
\noindent
\textbf{Sources of missing data.}
\label{sec:missing_data}
As in with any automated data collection campaign we are subject to missing data. It is important to understand and quantify sources of missing data.

First, some pages failed to load. In our experiments, page load timeout was set to 70 seconds.   If a page load fails, we do 2 more attempts, before we report an error and proceed with crawling. We found that 0.5\% of page loads timed out during our experiments.
Second, ad detection is performed automatically, hence we can have \emph{false positives} (page content is classified as an ad) and \emph{false negatives} (a real ad was classified as page content). Our classification algorithm is tuned towards reducing false positives rather than false negatives.
Third, in 1\% of cases we were not able to connect to the landing page URL for a given ad because they were calling a Javascript function. We were still able to use those ads in most of our analyses. However, we were unable to use those ads for landing page and advertiser analysis.

\section{Empirical Results}
\label{sec:results}
With over 875K ad impression and \adscount distinct ads harvested over a relatively short period of time with 340 distinct profiles, we believe that we can extract broad attributes of the Adscape.  We begin by considering the collection performance. 

\begin{figure}[t!]
\centering
  \psfig{file=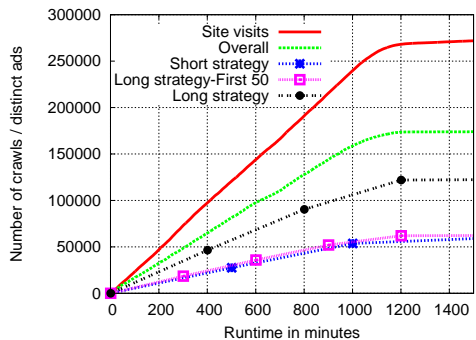, width=\figurewidthB, angle = -90}
\caption{Total number of unique ads collected.}
\vspace{-0.1in}
\label{fig:uniqueadsgrowth}
\end{figure}
\vspace{-0.1in}

\begin{figure}[t!]
\centering
  \psfig{file=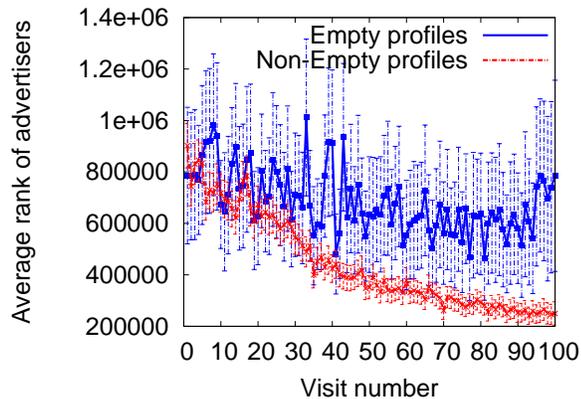, width=\figurewidthC, angle = -90}
\caption{Average traffic rank of advertisers, with 95\% confidence interval.}
\vspace{-0.1in}
\label{fig:advertiserrank}
\end{figure}
\vspace{-0.1in}

\medskip\noindent\textbf{Average Distinct Ad Arrival Rate.}
One objective of our study was to gather a large number of \emph{distinct ads}, where ads are distinct if the URL's of their creatives ({\em i.e.,} delivered images) are distinct. However, during harvesting we observed many duplicate ads.  This was not unexpected given the standard characteristics of ad campaigns and server configurations.  
Figure~\ref{fig:uniqueadsgrowth} shows the arrival rate of distinct ads at different stages of our two day crawling process. The $y$-axis shows the number of distinct ads collected and the $x$-axis shows runtime of the harvesting in minutes. The solid (red) line represents the total number of $(w, p)$-pairs visited and dash (green) line is the total number of distinct ads collected.  We can see that in the beginning of collection period the number of visits versus the number of unique ads grow at similar rates.  However, at some point the rate of new ads slows down.  This saturation point occurred approximately 1 day after we started the crawling process. Notice, that there is a plateau on the number of visits because our crawler instances finished their workload at different times.  

Interestingly, one can see that the {\em short} strategies harvest as many distinct ads as the first 50 visits of the {\em long} strategies (see Figure~\ref{fig:uniqueadsgrowth} ).  However, the long strategy still results in the largest number of ads  captured. Thus, as a design choice, if the goal is to gather a large number of distinct ads, neither strategy can be excluded.

\begin{figure}
  \centering
  \psfig{file=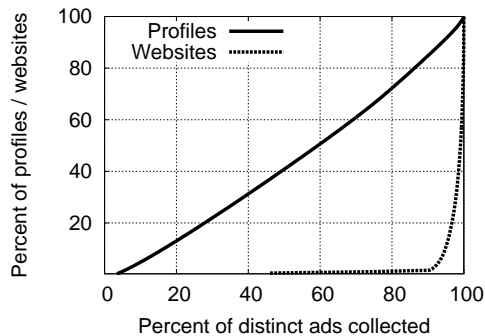, width=\figurewidthB, angle = -90}
\caption{Impact of websites and profiles on ads collection.}
\label{fig:impactofwebsitesprofiles}
\vspace{-0.1in}
\end{figure}
\vspace{-0.1in}

\medskip\noindent\textbf{Advertiser Rank Analysis.}
Next, we seek to understand the issue of ad targeting in greater detail.  Specifically, it may be the case that the value of an impression goes down as $w$ is visited repeatedly with a single $p$.  To test this hypothesis we have considered the average rank of the advertisers (acquired from Alexa) and the sequential number of the visit in the strategy.  We used only ads acquired by long strategies for this analysis. 58 of total \advertisercount advertisers do not have rank information available.  The average global rank for all available advertisers is ~900K. For each visit we compute the average rank of advertisers present across all pairs. Interestingly, Figure~\ref{fig:advertiserrank} shows that for pairs with non-empty $p$, the rank of the advertisers drops significantly over the course of visits. If $w$ is visited with an empty profile, the average rank of the advertiser stays virtually constant. 

Counter-intuitively, we see that the average rank of advertisers is decreasing, which implies that the advertisers shown in the later stage of the 100 visits are more popular.
In contrast, empty profiles do not see this trend of getting ads from more popular advertisers.  Instead, those advertisers' ranks remain roughly constant over 100 visits.

\medskip\noindent\textbf{Understanding Importance of $W$ and $P$.}
In Section~\ref{sec:focus_set} we observed that \textit{all} profiles from $P$ reached the focus set. The total number of distinct $w$ in the focus set is 180 from the initial 314. Does it mean that profiles play more significant role than websites?
To answer this question, we rank all websites by the number of distinct ads collected from each of them across profiles.  We do similarly for profiles.  Figure~\ref{fig:impactofwebsitesprofiles} shows that the number of distinct ads collected grows almost linearly with the number of profiles. However, for websites we observe the opposite effect.  Indeed, 2\% of all the websites used in data collection produced 90\% of all distinct ads observed!  This means that many of websites are redundant from the perspective of distinct ad collection, and one can concentrate on only few carefully chosen websites. However, it is not clear exactly how to select these sites.  We plan to address this issue in the future work.

\begin{figure}
  \centering
   \includegraphics[width=2.5in]{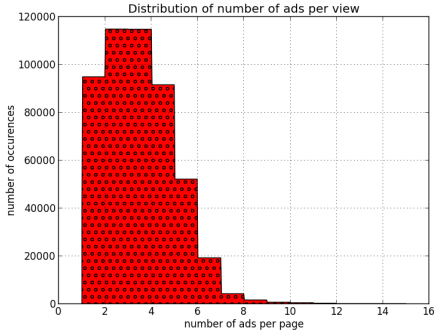}
\caption{Distribution of ad placements on web pages.}
\vspace{-0.15in}
\label{fig:adslots}
%\vspace{-0.15in}
\end{figure}

\medskip\noindent\textbf{Number of Ad Placements Per Page.}
Usually, a single webpage has more than one ad placement.  However, the number of placements can vary, and it is usually considered ``bad form" for a premium publisher to have too many placements (although there is no clear threshold).  The empirical distribution of the number of placements per page can be seen in Figure~\ref{fig:adslots}. The majority of pages have 2-3 placements, but there are some outliers that have as many as 16. The example of such an outlier is the blog \url{theberry.com}. 

\vspace{-0.1in}

\begin{figure}
  \centering
   \includegraphics[width=2.5in]{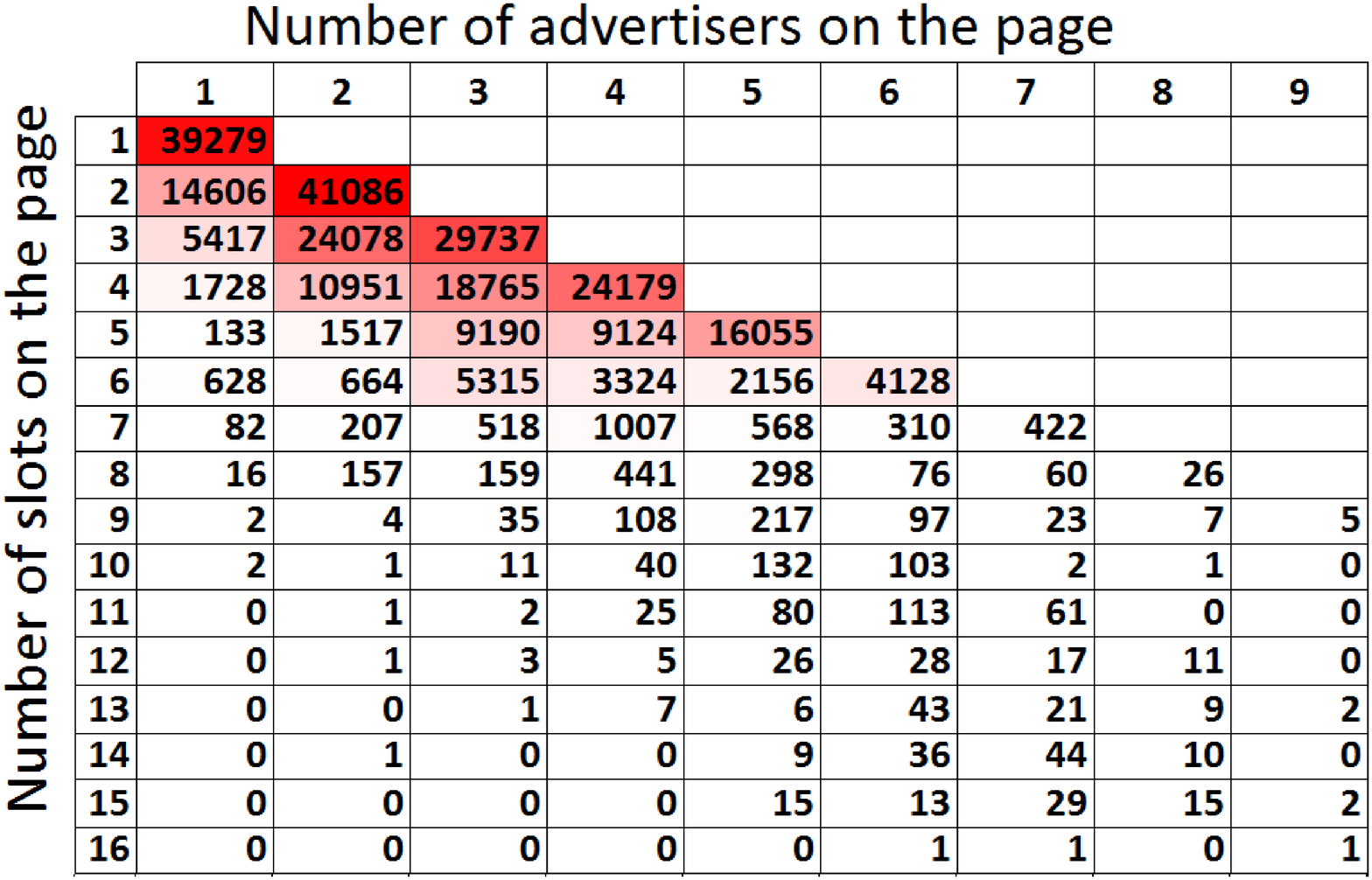}
\caption{Distribution of the number of advertisers on a single page.}
\vspace{-0.15in}
\label{fig:advertiserscountonsinglepage}
\vspace{-0.05in}
\end{figure}

\medskip\noindent\textbf{Number of Advertisers per Page.}
A question that has been considered formally (\eg \cite{feldman2010online}) is the optimal method for selling ad placements on a single page. Should all the display ads placements be allocated to a single advertiser exclusively? Or should they be shared among multiple advertisers? Figure~\ref{fig:advertiserscountonsinglepage} shows the empirical results for the number of advertisers placing ads per page. Each cell of the table is the number of the $(w,p)$-views that match the requirements. 

\vspace{-0.1in}

The figure show that exclusivity is indeed present, \eg there are 628 instances of pages with 6 placements that contained ads from a single advertiser.  However, multiple advertisers per page is the norm.  For instance, the vast majority (75\%) of 2 placement pages show ads that are delivered by different advertisers. A similar result also holds for 3 placements per pages, where 50\%  show ads delivered by 3 advertisers, 40\% for 2, and only 10\% for 1. 

\medskip\noindent\textbf{Frequency Capping.}
In this experiment we estimate and analyze frequency caps of ads in our dataset. 
Assume that some ads $a_i \in A$ have a frequency cap $f_i$. We cannot observe values of $f_i$. Let $f_{i}^e$ be \emph{empirical frequency} of ad $a_i$, $f_i^e$ is equal to maximum number of times ad $a_i$ was shown to any of observed pairs $(w, p)$. Let $c_i$ be the number of pairs that observed ad $a_i$ exactly $f_i^e$ times. 

We say that ad $a_i$ has an \emph{empirical frequency cap} if $c_i \ge C$, where $C$ is a manually picked constant. In our analysis, we set $C=5$. The intuition for this definition is as follows. If ad $a_i$ was shown to many pairs exactly $f_i^e$ times and there are no pairs that have seen ad $a_i$ more than $f_i^e$ times, then it is a strong indication that ad $a_i$ has $f_i = f_i^e$. We exclude the following ads from the pool:
{\em (1)} ads with $f_i^e=1$, \ie  ads that were shown at most once to any of the pairs; {\em (2)} ads that were shown to pair $(w,p)$ at almost every visit. 

We captured only 312 ads that satisfied our criteria for a frequency cap. 
However, the majority of ads have small frequency caps. 

\subsection{Profile Targeting Analysis}
\label{subsec:websitesdoprofiletargeting}
As discussed earlier, profiles play an important role in our experimental setup. Here we analyze the relationship between profile interest categories and ads that are served.

\medskip\noindent\textbf{Defining Targeted Ads.}
Intuitively, a targeted ad is shown to some profiles more frequently than to others. We use this intuition to define targeted ads. 
For each ad $a$ shown on a fixed website $w$, we count how many times it was shown to each of the profiles. 
This gives us empirical distribution $g_a(p)$ of ad $a$ over profiles $p$. If ad $a$ is not targeted based on user's profiles, it is fair to assume that the observed distribution $g_a$ should be close to uniform. To compare $g_a$ and uniform distribution we use Pearson's $\chi^2$ test.  We say that ad $a$ is \emph{targeted} if the resulting $p$-value is less than $0.05$.  Note that for some ads our sample size is too small, which makes it impossible to reject the hypothesis of uniform distribution, even if $g_a$ is not actually drawn from uniform distribution. Therefore, we might get false negatives \ie ads that are targeted but we falsely classify them as non-targeted.

\medskip\noindent\textbf{Measuring Targeted Ads.}

We calculate the fraction of targeted ads for each of the websites that are used in sufficient number of pairs (in our case, 10). Results are shown in Table~\ref{tab:targeting_profile}, column \emph{Profile}. The table shows that 50\% of the analyzed websites have at least 80\% of their ad inventory targeted at profiles. This  supports the observation that using a single profile to collect ads from many websites will not necessarily lead to a larger set of distinct ads.

\begin{table}
    \begin{tabular}{|l|l|l|l|}
    \hline
    \textbf{Websites}                        & \textbf{Profile} & \textbf{Gender}  & \textbf{Age}     \\ \hline \hline
    miamiherald.typepad.com         & 0.96    &    0.73 &    0.92 \\ \hline
       tech2.in.com                 &    0.92 &    0.61 &    0.84 \\ \hline
       chicago.cbslocal.com         &    0.86 &    0.68 &    0.70 \\ \hline
       moneycontrol.com             &    0.86 &    0.71 &    0.80 \\ \hline
       goal.com                     &    0.83 &    0.67 &    0.78 \\ \hline
       xda-developers.com           &    0.83 &    0.42 &    0.66 \\ \hline
       community.babycenter.com     &    0.83 &    0.36 &    0.70 \\ \hline
       celebritybabies.people.com   &    0.72 &    0.30 &    0.49 \\ \hline
       icanhas.cheezburger.com      &    0.71 &    0.26 &    0.34 \\ \hline
       women.webmd.com              &    0.71 &    0.38 &    0.49 \\ \hline
       lindaikeji.blogspot.com      &    0.69 &    0.48 &    0.52 \\ \hline
       ph.nba.com                   &    0.61 &    0.34 &    0.36 \\ \hline
       shechive.files.wordpress.com &    0.58 &    0.49 &    0.51 \\ \hline
    \end{tabular}
    \caption {Ratio of ads shown on websites whose distribution across the property does not follow uniform distribution with statistical significance.}
    \label{tab:targeting_profile}
    \vspace{-0.1in}
\end{table}
\vspace{-0.1in}

\medskip\noindent\textbf{Demographic Targeting.}
While our profiles are based on interest categories, some online services have also attributed demographic information to some of our profiles. For example, Google's ads settings associates our ``Pets" profile with a \emph{Female} from the 25-34 age group.   We fetch \emph{gender} and \emph{age groups} attribute values from profiles, and similar to interest targeting, we calculate for each website the fraction of ads targeted at age and gender. We excluded profiles that were not attributed, or attributed inconsistently, with demographic attributes.  We considered only websites that were used in at least 10 pairs.

Since there are only 5 age groups and 2 gender groups, our sample sizes per group are larger than in distributions per interest analysis.  Therefore, we classify ads into targeted/non-targeted groups with more confidence than in previous analysis.  Table~\ref{tab:targeting_profile} columns \emph{Gender} and \emph{Age} show that both age and gender are highly targeted attributes.  About half of the considered websites have more than 50\% of their ads targeted by gender. There is a similar targeting percentage by age. In summary, websites that show high level of targeting based on user interest profiles also show a high level of targeting based on gender and age group. Note, that here we quantify the portions of ads observed to bias towards some groups of age and gender.  We do not attempt to disentangle demographic-based targeting from interest-based targeting. We leave that as a task for future work.

\subsection{Advertiser Analysis}
\label{subsec:advertiserdistribution}
As noted above, during the course of our data collection we acquired ads from \advertisercount distinct advertisers.  We associate a distinct advertiser with the distinct domain of the landing page of an ad. We observe that more than 80\% of advertisers had no more than 100 ad impressions in our entire data corpus.  The top advertiser is \url{https://www.lmbinsurance.com}, which had a total of 94,437  impressions or about 10\% of all ad impressions. One of the websites we used in our crawling process shows consistently 2 ads linking to it on almost every visit by all profiles.  One possible explanation is, that \url{https://www.lmbinsurance.com} had an agreement or contract with that website. 

This phenomenon of a huge number of impressions from a single advertiser was not captured in while selecting the focus set, and cases like this makes the design of ad collection even more difficult.

Next, we endeavor to understand which ads target which profiles (or interest categories). For this analysis we identify the category of the advertiser of the ad. We say that an ad and it's advertiser have the same category.  We proceed by creating a correlation matrix between profiles' categories and ads' categories. 

\medskip\noindent\textbf{Advertiser Categories.}
To find a category for each advertiser, we appeal to multiple sources, since we did not find an authoritative source that would assign a category to all of the advertisers in our data set. 

We used site categorization services from Alexa \footnote{\url{http://www.alexa.com/topsites/category}} and WebPulse \footnote{\url{http://sitereview.bluecoat.com/sitereview.jsp}} from Blue Coat (which offers a web content filtering product). These two sources have different sets of categories for labeling sites, but most of the categories differences can be resolved manually. Table~\ref{tab:categorymapping} shows some samples of the category mappings made for Alexa and WebPulse.

\begin{table}
	\resizebox{\linewidth}{!}{%
    \begin{tabular}{|l|l|l|}
    \hline
    \textbf{Alexa} & \textbf{WebPulse} & \textbf{Mapped}   \\ 
    \hline 
    \hline
    Business/Automotive                     & Vehicles                                & Autos                                   \\ \hline
    Business/Hospitality/                     & \multirow{2}{*}{Restaurants} & \multirow{2}{*}{Restaurants} \\
    Restaurant Chains                          &                                             &                                             \\ \hline
    Business/Investing                         & Brokerage                             & Investing                              \\ \hline
    Reference/Education/                    & \multirow{2}{*}{Education}    & \multirow{2}{*}{Education}     \\
    Distance Learning/Online Courses &                                             &                                              \\ \hline
    Business/Real Estate                      & Real Estate                           & Real Estate                             \\ \hline
    Computers/Internet/                     & \multirow{3}{*}{Social Networking}  & \multirow{3}{*}{Social Networking} \\
    On the Web/Online Communities/ &                                             &                                               \\
    Social Networking                          &                                             &                                               \\ \hline
    Recreation/Travel                          & Travel                                   & Travel                                    \\ \hline
    \end{tabular} }

    \caption{Sample category mappings for Alexa and WebPulse.}
    \label{tab:categorymapping}
\vspace{-0.1in}
\end{table}
\vspace{-0.1in}
 
\begin{figure}
   \psfig{file=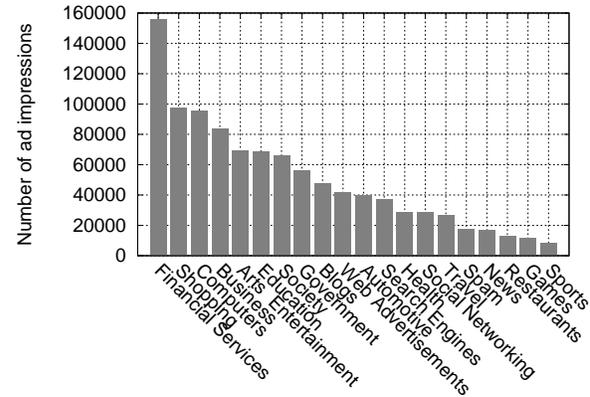, width=\figurewidthC, angle=-90}
\caption{Distribution of impressions over categories.}
\vspace{-0.2in}
\label{fig:adsdistribbycat}
\end{figure}

We categorized \alexacount of \advertisercount advertisers using Alexa, and used WebPulse for the rest.   Categorized impressions fall into more than 550 different detailed categories.  Figure~\ref{fig:adsdistribbycat} shows ad distribution for the top 20 root level categories.  The top advertiser category is \emph{Financial Services}, which also contains the top advertiser.  Some other major categories are \emph{Shopping}, \emph{Computers}, {Business}, \emph{Arts \& Entertainment} and \emph{Education}. We have also collected ads from 17 advertisers that have been categorized as \emph{Spam}. 

\vspace{-0.1in}

\begin{figure*}
 \includegraphics[width=7in]{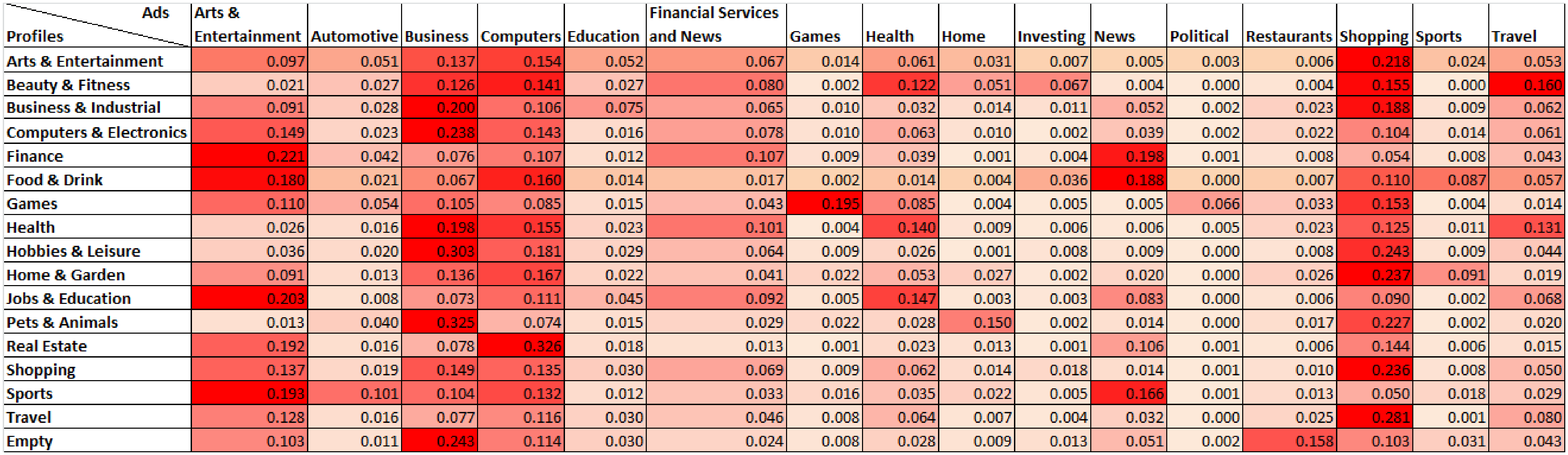}
\caption{Heat map for profile and ads categories. Row-wise normalized based on number of ad impressions.}
\vspace{-0.15in}
\label{fig:heatmap_profile}
\end{figure*}
\vspace{-0.1in}

\medskip\noindent\textbf{Profiles and Ad Categories.}
It is natural to expect that advertisers target different types of users.  In this experiment, we are looking for the relation between interest categories of $p$'s and categories $c$'s of ads. (For each ad impression, we consider the category $c$ of the corresponding advertiser.)  We attribute the appearance of $c$ to all the profiles $p$'s that observed the ad.   We normalize this by the total number of $p$'s that saw the ad, to discount for the fact that many different profiles saw it, possibly without being the focus of targeting. After processing all ad impressions, for each interest categories of $p$, we produce its impression distributions over different ad categories. 

Figure~\ref{fig:heatmap_profile} shows the result (for space considerations, we only show top level profiles interest and ad categories).  This heat-map is row-wise normalized, such that it is easy to tell which ad categories that an interest category is likely to see uniquely.  Several characteristics are immediately evident:  
\begin{inparaenum}[(1)]
\item Some $p$'s and $c$'s exhibit  high correlation, for example \emph{Games}, \emph{Health} and \emph{Shopping}; 
\item Some profiles are targeted by related categories, for example \emph{Beauty \& Fitness} profile is highly targeted by \emph{Shopping} and \emph{Travel}, and \emph{Pets \& Animals} profile is highly targeted by \emph{Home} related ads; 
\item Ads from categories \emph{Arts \& Entertainment}, \emph{Business}, \emph{Computers} and \emph{Shopping} are less targeted;
\item Interestingly, the empty profile is highly targeted by \emph{Restaurants} ads, which are rarely seen by any of the other profiles. 
\end{inparaenum}

\section{Related Work}
\label{sec:related}

Our ad crawling capability is most directly related to standard \emph{web crawling}, which is widely used to gather content for search engines and a host of other applications.  Early web crawlers emerged nearly two decades ago including  WebCrawler~\cite{pinkerton1994finding}, World Wide Web Worm~\cite{mcbryan1994genvl} and RBSE~\cite{eichmann1994rbse}.  Googlebot~\footnote{\url{http://support.google.com/webmasters/bin/answer.py-?hl=en&answer=182072}}
and Bingbot~\footnote{\url{http://www.bing.com/blogs/site_blogs/b/webmaster/-archive/2010/09/03/bingbot-is-coming-to-town.aspx}} are two of the most prominent examples of modern web crawlers.  The on-going challenges in content crawling include the ever-increasing number of websites, increasing use of dynamic content, and the tension between crawling frequency for information freshness and demand on Internet resources.  Examples of studies that consider these problems include~\cite{Arasu03,Broder03,Chang06,Olsten10}.

Pandey and Olsten consider ``user centric'' web crawling in~\cite{Pandey05}.  Their focus is on scheduling web crawls to specific pages in order to maintain the most up to date versions in search engine repositories.  At the highest level, elements of our ad crawling system have similar objectives.  More recently, Liu {\em et al.} consider the problem of using hints from user browser histories to organize URL lists for crawling~\cite{Liu11}.  We are aware of no prior work that builds and employs user profiles for ad crawling in the way that we do.

The task of crawling ads differs significantly from web crawling in a number of ways.  These include but are not limited to the fact that {\em (i)} different ads can be shown to a user on each page reload, {\em (ii)} ads are delivered based on information beyond page context, {\em (iii)} display ads must be differentiated from other visual elements on a page.  We are not aware of any work describing methods for crawling display ads.  However, there are number of plugins that allow users to filter out display ads.  AdBlock~\footnote{\url{http://adblockplus.org}} is one of most widely used.  

Several studies consider the problems associated with privacy and online advertising.  Castelluccia {\em et al.}~\cite{castelluccia2012betrayed} demonstrated that one can reverse engineer users' profiles by looking at targeted ads displayed to her and making inferences about the target interests revealed in ads. In their work, they focused on root level categories of tree of profile interests that can be found on \url{www.google.com/ads/preferences/}.   Roesner {\em et al.}~\cite{Roesner2012detectthirdparty} present a taxonomy of different trackers {\em i.e.,} in-site, cross site, cookie sharing, and social media trackers.  In experiments, authors simulated users using AOL query search logs, and considered how prevalent tracking is in this dataset and propose an extension that helps to protect user privacy.  Finally, the study by Guha {\em et al.}~\cite{guha2010challenges}, which described challenges in measuring online advertising systems, informs our work.  However, their primary focus is on privacy issues, while ours is on broader Adscape characterization.

\section{conclusions}
\label{sec:summery}

In this paper we describe a novel study of online display advertising {\em i.e.,} the {\em Internet Adscape}.  The goals of our work are to develop a general understanding of the characteristics and dynamics of online display advertising and to gain insights on the targeting mechanisms that are used by ad serving entities.  Our work beings by developing methods and tools for gathering display ads from a large number of web sites.  Central to our ad crawling method is the use of {\em user profiles}, which have an influence on the particular ads that are served.  We developed a scalable crawling infrastructure based on the Firefox browser that distinguishes display ads from other images and gathers all relevant information (creatives, landing pages, etc.).  We use this infrastructure to crawl a set of over \websitecount English-language web sites using \profilecount different profiles.  The result of this crawl is a set of over 175K unique ads that are the basis of our evaluation.

We analyze our ad data corpus from three different perspectives.  First, we consider the general use of targeted advertising and find that the majority of sites that we visited employing targeting mechanisms on over 80\% of their display ads. However, we also find that many sites deliver ads to all users regardless of profile.  This is most likely to be explained by a lack of targeting specificity by an advertiser.   Next, we evaluate the population of Marketers who are engaged in online display advertising.  We identify over 3.7K unique Marketers from diverse business segments. Finally, we consider the details of display ads themselves in terms of content and consistency with user profiles. We find that there is generally an alignment between delivered ads and user profile. 

Our on-going activities include continuing and expanding our data gathering efforts.  By expanding our data set, in future work we will conduct longitudinal analyses to enhance our understanding of ad delivery mechanisms and ad campaign dynamics.  We also plan to expand our focus beyond display advertising to include video advertising.  Finally, we plan to drill down in greater detail on the mechanisms for ad targeting and build models that we can use to develop improved methods.

\section*{Acknowledgements} 

This work was supported in part by NSF grant CNS-0905186, ARL/ARO grant W911NF1110227, the DHS PREDICT Project, and the NSF Grant 1101677: ICES: Auctions and Optimizations in Ad Exchanges.  Any opinions, findings, conclusions or other recommendations expressed in this material are those of the authors and do not necessarily reflect the views of the NSF, ARO or DHS.

\bibliographystyle{abbrv}
\bibliography{lib} 

\end{document}